\documentclass[report,11pt]{article}

\usepackage{tikz-cd}
\usepackage{scrextend}
\usepackage{amsmath,amssymb,epsfig,bbm}
\usepackage{MnSymbol}
\usepackage{slashed}
\usepackage{amsthm}

\usepackage{float}

\usepackage{enumerate}

\usepackage{color}
\usepackage{mathrsfs}

\usepackage{dsfont}

\definecolor{co}{cmyk}{0,0.7,0.3,0}
\definecolor{darkgreen}{cmyk}{1,0,1,.2}
\definecolor{m}{rgb}{1,0.1,1}

\newcommand{\be}{\begin{equation}}
\newcommand{\ba}{\begin{eqnarray}}
\newcommand{\ea}{\end{eqnarray}}
\newcommand{\nn}{\nonumber}

\def\a{\alpha}
\def\b{\beta}

\def\m{\mu}
\def\n{\nu}
\def\oo{\omega}

\def\OO{\Omega}

\def\ca{{\cal A}}
\def\cb{{\cal B}}

\def\cf{{\cal F}}
\def\cg{{\cal G}}
\def\ch{{\cal H}}

\def\co{{\cal O}}

\makeatletter
\newcommand{\eqnum}{\refstepcounter{equation}\textup{\tagform@{\theequation}}}
\makeatother

\newcommand{\pa}{\partial}

\newtheorem{thm}{Theorem}[subsection]

\newtheorem{definition}[thm]{Definition}

\newtheorem*{definition*}{Definition}

\fontfamily{yfrak}

\newcommand{\xdownarrow}[1]{%
  {\left\downarrow\vbox to #1{}\right.\kern-\nulldelimiterspace}
}

\DeclareRobustCommand{\looongrightarrow}{%
  \DOTSB\relbar\joinrel\relbar\joinrel\relbar\joinrel\rightarrow
}

\begin{document}

\vskip 25mm

\begin{center}

{\large\bfseries

On the emergence of an almost-commutative spectral triple from a geometric construction on a configuration space

}

\vskip 6ex

Johannes \textsc{Aastrup}$^{a}$\footnote{email: \texttt{aastrup@math.uni-hannover.de}} \&
Jesper M\o ller \textsc{Grimstrup}$^{b}$\footnote{email: \texttt{jesper.grimstrup@gmail.com}}\\ 
\vskip 3ex

$^{a}\,$\textit{Mathematisches Institut, Universit\"at Hannover, \\ Welfengarten 1, 
D-30167 Hannover, Germany.}
\\[3ex]
$^{b}\,$\textit{Copenhagen, Denmark.}
\\[3ex]

{\footnotesize\it This work is financially supported by entrepreneurs \\Kasper Bloch Gevaldig, Denmark, and Jeff Cordova, USA.}

\end{center}

\vskip 3ex

\begin{abstract}

\vspace{0.5cm}

We show that the structure of an almost-commutative spectral triple emerges in a semi-classical limit from a geometric construction on a configuration space of gauge connections. The geometric construction resembles that of a  spectral triple with a Dirac operator on the configuration space that interacts with the so-called $\mathbf{HD}$-algebra, which is an algebra of operator-valued functions on the configuration space, and which is generated by parallel-transports along flows of vector-fields on the underlying manifold. In a semi-classical limit the $\mathbf{HD}$-algebra produces an almost-commutative algebra where the finite factor depends on the representation of the $\mathbf{HD}$-algebra and on the point in the configuration space over which the semi-classical state is localized. Interestingly, we find that the Hilbert space, in which the almost-commutative algebra acts, comes with a double fermionic structure that resembles the fermionic doubling found in the noncommutative formulation of the standard model. Finally, the emerging almost-commutative algebra interacts with a spatial Dirac operator that emerges in the semi-classical limit. This interaction involves both factors of the almost-commutative algebra.

\end{abstract}

\newpage
\tableofcontents

\section{Introduction}

One of the most interesting developments in theoretical high-energy physics since the discovery of the standard model of particle physics is the realisation that it is possible to formulate the standard model coupled to general relativity as a single gravitational theory using noncommutative geometry \cite{Connes:1990qp}-\cite{Chamseddine:2007hz}. 
This discovery entails on the one hand novel empirical predictions \cite{Chamseddine:2006ep} while it on the other hand ties together some of the most important unsolved problems in the field: the origin of the algebraic structure of the standard model, the rigorous formulation of quantum field theory, and the reconcilement of general relativity with quantum theory.

The spectral formulation of the standard model builds on the reconstruction theorem \cite{Connes:2008vs} that states that Riemannian spin-geometry of a compact manifold $M$ has an equivalent formulation based on the spectral triple
\begin{equation}
(C^\infty(M),L^2(M,S),D_M)
\label{eeet}
\end{equation}
where the $D_M$ is the Dirac operator on $M$.
The spectral triple $(\cb,\ch,D)$ that corresponds to the standard model is then the tensor product between the spectral triple in (\ref{eeet}) and a finite-dimensional spectral triple 
 $
(\cb_F, \ch_F,D_F)
$
where 
$$\cb_F=\mathbb{C}\oplus \mathbb{H}\oplus M_3(\mathbb{C})$$ is a matrix algebra that matches the gauge structure of the standard model, $D_F$ is a finite-dimensional Dirac operator built from the Yukawa coupling
matrix \cite{Connes:1996gi}, and $\ch_F$ is a Hilbert space that encodes the particle content of the standard model. The geometry of $\cb$ is then given by
\begin{equation}
\cb = C^\infty \otimes \cb_F, \quad \ch = L^2(M,S)\otimes \ch_F,\quad  D = D_M \otimes 1 + \gamma_5\otimes D_F.
\label{ccz}
\end{equation}
A classification of finite-dimensional algebras that satisfy the axioms of noncommutative geometry as well as an additional assumption about its KO-dimension equaling 2 mod 8 was carried out in \cite{Chamseddine:2007hz}. This classification lead to a sequence of finite-dimensional algebras 
\begin{equation}
M_k(\mathbb{C})\oplus M_k(\mathbb{C})
\longrightarrow 
M_4(\mathbb{C})\oplus M_2(\mathbb{H})
\longrightarrow
\cb_F,
\label{manon}
\end{equation}
where $\mathbb{H}$ are the quaternions, that singles out $\cb_F$ as unique within the axiomatic system of noncommutative geometry.

These remarkable results raise two fundamental questions
\begin{enumerate}
    \item 
    Where does the almost commutative spectral triple behind the standard model originate from?
    \item 
    How does quantum field theory fit into this framework?
\end{enumerate}
The second question stems from the fact that this new formulation of the standard model does not include quantum field theory as a primary ingredient. Rather, the spectral action principle \cite{Chamseddine:1996zu} is used to derive the Lagrangian of the standard model coupled to general relativity from the spectral triple, and the standard model is then subsequently quantized using perturbative quantum field theory and renormalisation group theory while gravity is left untouched. However, if this new formulation of the standard model is to be fundamental -- and we certainly believe that this is the case -- then surely quantum field theory must in some fashion be included as a primary ingredient? Hence the question.

In a long series of papers we have proposed that the answer to these two questions should be sought in a geometrical construction on a configuration space of gauge connections \cite{Aastrup:2005yk}-\cite{Aastrup:2024iqe}.
 In previous publications we have already shown that a spectral triple-like construction over a configuration space of gauge connections gives rise to many of the key building blocks of contemporary high-energy physics: the canonical commutation and anti-commutation relations of bosonic and fermionic quantum field theory, the Hamilton operator of a Yang-Mills quantum field theory coupled to a fermionic sector, and also elements of general relativity. 

In this paper we show that the general structure of a Hamilton formulation of an almost-commutative spectral triple also emerges from this construction in a semi-classical limit. Specifically, in the special case where the configuration space consist of  $U(2)$ gauge connections and under the assumption that the semi-classical limit provides us with a metric on the underlying manifold -- something we have previously argued should be the case \cite{Aastrup:2023jfw} -- then the fermionic Hamilton operator can be reformulated in terms of fermions with half-integer spin, and the emerging almost-commutative geometry is found to involve a finite-dimensional factor that is a sub-algebra of $M_8(\mathbb{C})$. The specifics of the sub-algebra depends on the details of the semi-classical limit as well as a choice of representation of the $\mathbf{HD}$-algebra.

To explain how an almost-commutative algebra emerges from a geometrical construction over a configuration space we first note that the algebra used in this construction is the so-called $\mathbf{HD}$-algebra \cite{Aastrup:2012vq}, which is generated by parallel transports along flows of vector fields on a three-dimensional manifold. The $\mathbf{HD}$-algebra consist of operator-valued functions on the underlying configuration space and what we find is that in a semi-classical limit the holonomy transforms along closed loops give rise to an almost commutative algebra that consist of smooth functions on the underlying manifold together with a matrix algebra that depends on the representation of the $\mathbf{HD}$-algebra. 

A central feature of Chamseddine and Connes' formulation of the standard model has to do with a certain doubling of fermionic degrees of freedom in the sense that they are counted twice in the Hilbert space $\ch$: once in $L^2(M,S)$, and once in $\ch_F$  \cite{Lizzi:1996vr,Gracia-Bondia:1997adl}. The problems that this over-counting leads to was solved by setting the KO-dimension of the noncommutative geometry to equal 2 mod 8 \cite{Barrett:2006qq,Chamseddine:2006ep}. Interestingly, we find the same type of fermion doubling in our geometrical construction over the configuration space. In the semi-classical limit the representation of the $\mathbf{HD}$-algebra splits into two factors, both of which carries fermionic degrees of freedom. The one factor is related to the underlying manifold $M$, the other is related to the finite-dimensional factor. \\

This paper is organized as follows: First we review previous results to set the stage for our analysis: in section 2 we go through the construction of the $\mathbf{HD}$-algebra, in section 3 we introduce the geometrical construction over the corresponding configuration space and derive the Yang-Mills and Dirac Hamiltonians. In section 4 we then reformulate the fermionic Hamiltonian in terms of spinors, and in section 5 we introduce a Hilbert space representation and explain how an almost-commutive algebra emerges in a semi-classical limit. We also show how this almost-commutative algebra interacts with the Dirac Hamiltonian. We end with a discussion in section 6.

\section{The $\mathbf{HD}$ algebra}

We begin by introducing the
$\mathbf{HD}$-algebra. The following is based on \cite{Aastrup:2012vq}. 
Let $M$ be a three-dimensional manifold $M$ and let\footnote{We have previously defined the $\mathbf{HD}$-algebra either with a general Lie-group or with $SU(2)$. However, for reasons that will become clear later we choose$U(2)$ for the remainder of this paper. } $\mathfrak{u}(2)$ be the Lie-algebra of $U(2)$. We denote by $V$ a bundle over $M$ in which a $\mathfrak{u}(2)$-valued connection $A$ acts. Let $X$ be a vector-field on $M$ and $t\to \exp_t(X)$ the corresponding flow. Given $x\in M$ let $\gamma$ be the curve  
$$\gamma (t)=\exp_{t} (X) (x) $$
running from $x$ to $\exp_1 (X)(x)\in M$. We define the operator 
$$e^X_A :L^2 (M , V) \to L^2 (M ,  V )$$
in the following way:
we consider an element $\xi \in L^2 (M ,  V)$ as a function with values in $V$, and define 
\begin{equation}
  (e^X_A \xi )(\exp_1(X) (x))=  ((\Delta \exp_1) (x))  \hbox{Hol}(\gamma, A) \xi (x)   ,
  \label{chopin1}
 \end{equation}
where $\hbox{Hol}(\gamma, A)$ denotes the holonomy of $A$ along $\gamma$ and where $\Delta$ is a factor that secures that $e^X$ is a unitary operator (see \cite{Aastrup:2012vn} for details). 
This gives us an operator valued function on the configuration space $\ca$ of $\mathfrak{u}(2)$-valued connections defined via 
\begin{equation}
\ca \ni A \to e^X_A  . 
\nn
\end{equation}
We denote this function $e^X$ and call it a holonomy-diffeomorphisms. 
Denote by $\mathscr{F} (\ca , \mathbb{B} (L^2(M, V) ))$ the bounded operator valued functions over $\ca$. This forms a $C^*$-algebra with the norm
$$\| \Psi \| =  \sup_{A \in \ca} \{\|  \Psi (A )\| \}, \quad \Psi \in  \mathscr{F} (\ca , \mathbb{B} (L^2(M, V )) ). $$ 
    The following definition was first given in \cite{Aastrup:2012vq}:

\begin{definition}
Let 
$$C =   \hbox{span} \{ e^X | \ X \hbox{ vector field on  } M\}  . $$
The holonomy-diffeomorphism algebra $\mathbf{H D} (M,V,\ca)   $ is defined to be the $C^*$-sub-algebra of  $\mathscr{F} (\ca , \cb (L^2(M,V )) )$ generated by $C$.
We will often denote $\mathbf{H D} (M,S,\ca)   $ by  $\mathbf{H D}  $ when it is clear which $M$,$S$, and $\ca$ are meant.
\end{definition}

In \cite{Aastrup:2012vn} we proved that  $\mathbf{H D} (M,S,\ca)   $ is independent of the metric $g$ on $M$. For further details on the $\mathbf{H D} (M,S,\ca)   $ algebra see \cite{Aastrup:2012vq,Aastrup:2012vn,Aastrup:2014ppa}. 


\section{A spectral triple-like construction on a configuration space}

Once we have the configuration space $\ca$ and the $\mathbf{H D}$-algebra of operator-valued functions on $\ca$ we can consider its geometry. In particular, we can construct what resembles a spectral triple \cite{Connes:1996gi} over $\ca$, i.e. a Dirac operator on $\ca$ that interacts with the $\mathbf{H D}$-algebra, as well as a Hilbert space $\ch$ over $\ca$.

\subsection{Gauge fixing}

Due to the gauge symmetry we shall in part work with a gauge-fixing $\cf$ of $\ca$. This means that we require that for each $A \in \ca$ there is exactly one $g\in \cg$ with $g(A) \in \cf $, where $\cg$ is the space of gauge transformations. The correct treatment of the gauge fixing involves a BRST procedure adapted to the present setup. This was done in \cite{Aastrup:2023jfw} and we refer the reader to that paper for details. 

In the following we will give a  brief outline of how a Dirac operator can be formulated on $\cf$ and how a Hamiltonian formulation of a Yang-Mills-Dirac quantum field theory can be obtained from such a construction. For further details, we refer the reader to \cite{Aastrup:2023jfw}-\cite{Aastrup:2024iqe}\cite{Aastrup:2024xxl}.

\subsection{A metric on the configuration space}

In order to construct a Dirac operator on $\cf$ we first need to introduce a metric on $\ca$. To do this we first note that if we choose an element $A_0\in\ca$ then we can write any connection $A\in \ca$ as
$$
A = A_0 + \oo
$$
where $\oo\in \OO^1(M,\mathfrak{u}(2))$ is a one-form that takes values in $\mathfrak{u}(2)$. We can use this to write the tangent space of $\ca$ in $A_0$ as
$$
T_{A_0}\ca = \OO^1(M,\mathfrak{u}(2)),
$$
which means that $T\ca = \ca \times \OO^1(M,\mathfrak{u}(2))$ (for details see \cite{Aastrup:2020jcf}). 
Next, we assume that a gauge-covariant metric on $\ca$ exists that 
\begin{enumerate}
\item  is fibered over $\ca$, i.e. that it is of the type
\begin{equation}
\ca\ni A \to \langle \cdot , \cdot \rangle_{A} ,
\label{innerspinor}
\end{equation}
where $\langle \cdot \vert \cdot \rangle_{A}$ is an inner product on $\OO^1(M,\mathfrak{u}(2))$, 
\item permits the construction of the Dirac operator on $\cf$ and the Hilbert space $L^2(\cf)$ that we will discuss in the following. In \cite{Aastrup:2023jfw} we proved that a metric of this type can be constructed. 
\end{enumerate}

In \cite{Aastrup:2024xxl} we 
constructed a map between the space $\OO^1(M,\mathfrak{u}(2))$ and the space $\OO^1(M,S\oplus S)$, where $S$ is the spin-bundle over $M$. The motivation for constructing this map was to introduce half-integer spin into the construction and thus avoid having fermions with integer spin, which is in conflict with the spin-statistics theorem.
To see how this map is constructed we first note that there exist a map 
$$ P:\Omega^1 (M ,\mathfrak{u}(2) )  \times C^\infty (M,S\oplus S) \to  \Omega^1 (M ,S\oplus S ),$$
since $\mathfrak{u}(2)$ acts on $S$. Next, we choose $(\varphi_1,\varphi_2) \in C^\infty(M,S \oplus S)$ and construct the linear map
\begin{equation}
\chi_{(\varphi_1,\varphi_2) }:\Omega^1 (M,\mathfrak{u}(2))\to \Omega^1 (M,S\oplus S)
\label{chiii}
\end{equation}
given by 
$$
\chi_{(\varphi_1,\varphi_2) } (\omega) =P(\omega,(\varphi_1,\varphi_2)  ).
$$

The space $\Omega^1 (M, \mathfrak{u}(2))$ is a priory a real vector space whereas $\Omega^1 (M,S\oplus S)$ is a complex vector space. Note however that if consider 
\begin{equation}
\chi_{(\varphi_1, \varphi_2)}\otimes_{\mathbb{R}} \mathbb{C} :\Omega^1 (M, \mathfrak{u}(2)) \otimes  _{\mathbb{R}} \mathbb{C} \to  \Omega^1 (M,S\oplus S) 
\label{alien}
\end{equation}
then this is a   linear isomorphism if $(\varphi_1,\varphi_2)$ is linear independent in each point on $M$. 

The metric $\ca \ni A \to \langle \cdot , \cdot \rangle_A$ gives a Hilbert space structure on 
$$L^2 (\cf , \bigwedge^*  \Omega^1 (M,\mathfrak{u}(2))   ) .$$
Note here that $\bigwedge^*  \Omega^1 (M,\mathfrak{u}(2))$ is here considered a complex vector space. 

We can use $\chi_{(\varphi_1, \varphi_2)}\otimes_{\mathbb{R}} \mathbb{C}$  to transport the metric from $\Omega^1 (M,\mathfrak{u}(2))   $ to $\Omega^1 (M,S\oplus S)$, which we denote $\langle \cdot , \cdot \rangle_S$, and thus use $\chi_{(\varphi_1, \varphi_2)}\otimes_{\mathbb{R}} \mathbb{C}$ to identify  $\bigwedge^*  \Omega^1 (M,\mathfrak{u}(2))$ with $\bigwedge^*  \Omega^1 (M,S \oplus S)$.

In \cite{Aastrup:2024xxl} we found that the metric $\langle \cdot , \cdot \rangle_S$ is invariant under a transformation of the two spinors $\varphi_1$ and $\varphi_2$, i.e. $\varphi'_1= N\varphi_1$ and $\varphi'_2= N\varphi_2$ where $N$ is a unitary matrix, whenever $N$ lies in the kernel of the Laplace operator $\Delta^A$. Since the Laplace operator depends on $A$ this implies that independence can only be achieved in a semi-classical limit centered over a classical connection $A_0$.

In this paper we shall use the map $\chi_{(\varphi_1, \varphi_2)}$ in two separate circumstances, one as a map between the spaces $\OO^1(M,\mathfrak{u}(2))$ and the space $\OO^1(M,S\oplus S)$, and one as a map between $L^2(M,\mathfrak{su}(2), V)$ and $L^2(M,S\oplus S, V)$ where $V$ is a vector-space.



\subsection{The Dirac and Hamilton operators}

We are now going to construct a Dirac operator on $\cf$. Since we have a metric $\langle \cdot,\cdot \rangle_A$ on $\ca$ we can divide the tangent space of $\ca$ into orthogonal parts:
\begin{equation}
T\ca = T\cf \oplus (T\cf)^\perp.
\label{tsktsk}
\end{equation}
We will denote by $\{\xi_i\}$  a basis of $\OO^1(M,\mathfrak{u}(2))$ that is orthonormal with respect to $\langle \cdot,\cdot \rangle_A$ and which can be divided into orthogonal parts according to the direct sum in (\ref{tsktsk}). 

With this we write down the Dirac operator 
\begin{equation}
D_\cf =
\left(
\begin{array}{cc}
D_1 & 0 \\
0 & D_2
\end{array}
\right)
\label{holger}
\end{equation}
with 
\begin{equation}
D_1 = \sum_i \bar{c}(\xi_i) \nabla_{\xi_i},\quad
D_2= \sum_i \bar{c}(\mathrm{i}\xi_i) \nabla_{\xi_i},
\label{Finn}
\end{equation}
where we only sum over vectors $\xi_i$ that belong to $T\cf$. Also, $\nabla_{\xi_i}$ in (\ref{Finn}) is the covariant derivative in the direction of $\xi_i$ given by the metric on $\ca$. 
Finally, the Clifford multiplication operators $\bar{c}(\xi)$ and $c(\xi)$ in (\ref{Finn}) are given by
\begin{eqnarray}
  c(\xi) &=& \mbox{ext}(\xi) + \mbox{int}(\xi),
\nn\\
 \bar{c}(\xi) &=& \mbox{ext}(\xi) - \mbox{int}(\xi) 
\nn%
\end{eqnarray}
where $\mbox{ext}(\xi)$ and $ \mbox{int}(\xi) $ are the operators of exterior and interior multiplication in $\bigwedge^* \OO^1(M,\mathfrak{u}(2))$ \cite{Aastrup:2024xxl}.

In \cite{Aastrup:2024iqe} we analysed the unitary fluctuations of the Dirac operator $D_\cf$ and found that the square of a Dirac operator $\tilde{D}_\cf$, which is obtained by modifying $D_\cf$ with a type of 'twisted' unitary fluctuation, gives rise to the Hamiltonian for a Yang-Mills quantum field theory coupled to a fermionic sector. The operator 
$\tilde{D}_\cf$ is obtained from $D_\cf$ in the following way 
\begin{equation}
 \tilde{D}_\cf = D_\cf + \gamma u [D_\cf,u^{-1}] \gamma^{-1},
\label{dune}
\end{equation}
with 
$$
u = 
\left(
\begin{array}{cc}
\exp \left( \mathrm{i} CS(A) \right) & 0 \\
0 & \exp \left( -\mathrm{i} CS(A) \right)
\end{array}
\right),\qquad 
\gamma =
\left(
\begin{array}{cc}
0 & 1 \\
1 & 0
\end{array}
\right),
$$
where
\begin{eqnarray}
CS(A) &=&  
\int_M \mbox{Tr} \left( {A}\wedge d{A} + \frac{2}{3} {A}\wedge {A} \wedge {A}\  \right) 
\nn%
\label{CS}
\end{eqnarray}
is the Chern-Simons term.

In \cite{Aastrup:2024iqe,Aastrup:2024xxl,Aastrup:2024ytt} we found that if we perform the unitary fluctuation (\ref{dune}) without the 'twist' given by the $\gamma$-operator, i.e. if we instead use\footnote{The Dirac operator used in \cite{Aastrup:2024xxl} was slightly different from the one used here. The difference is the definition of $D_2$, which in \cite{Aastrup:2024xxl} involved a real structure, whereas it is here involves the complex $i$. This difference is, however, not important for the point that we wish to make here, i.e. a unitary fluctuation of the Dirac operator used in this paper would yield the same result as found in \cite{Aastrup:2024xxl}. } $\tilde{D}_\cf= D_\cf +  u [D_\cf,u^{-1}] $, then we obtain the self-dual and anti-self-dual sectors of a Yang-Mills theory as well as a spectral invariant but without a fermionic sector.

For the Dirac operator $\tilde{D}_\cf$ defined in (\ref{dune}) a straightforward computation \cite{Aastrup:2024iqe} gives 
\begin{equation}
    \tilde{D}^2_\cf = 
  \left(
\begin{array}{cc}
H_{\mbox{\tiny YM}} + H_{\mbox{\tiny fermionic}}  & 0 \\
0 & H_{\mbox{\tiny YM}} + H_{\mbox{\tiny fermionic}}
\end{array}
\right) ,
\nn
\end{equation}
with
\begin{eqnarray}
H_{\mbox{\tiny YM}} &=& \sum_i\left(  -\left( \nabla_{\xi_i} \right)^2 + \left( [\nabla_{\xi_i}, CS(A)] \right)^2 \right),
\nn\\
 H_{\mbox{\tiny fermionic}} &=& \mathrm{i} \left\{ D_1, [D_2, CS(A)]  \right\} .
\nn\label{ddd}
\end{eqnarray}
In \cite{Aastrup:2023jfw} we identified the first term $H_{\mbox{\tiny YM}}$ as the Hamiltonian of a Yang-Mills quantum field theory (see also \cite{Aastrup:2020jcf}). Furthermore, in \cite{Aastrup:2024ytt} we rewrote the second term $H_{\mbox{\tiny fermionic}}$ as
\begin{equation}
H_{\mbox{\tiny fermionic}} = 2 \int_M \mbox{Tr}\left( \Phi  \nabla^A \Phi^\dagger -  \Phi^\dagger  \nabla^A \Phi  \right)  + \Xi ,
\label{kam}
\end{equation}
where we used the relation $\bar{c}(\mathrm{i}\xi) = \mathrm{i} c(\xi)$ together with
\begin{equation}
 \frac{\pa^2 CS(A)}{\pa x_{i } \pa x_{j }} 
 =    \int_M \mbox{Tr} \left(  \xi_{i  }  \wedge \nabla^A  \xi_{j} \right) + \int_M \mbox{Tr} \left(  \xi_{j  }  \wedge \nabla^A  \xi_{i} \right) ,
\label{second}
\end{equation}
where $\nabla^A$ is the covariant derivative and where we defined the operator-valued fermionic fields
\begin{equation}
\Phi(x) = \sum_i \xi_i(x) \mbox{int}(\xi_i), \quad 
\Phi^\dagger (x) = \sum_i \xi_i(x) \mbox{ext}(\xi_i).
\label{llutz}
\end{equation}
The additional term $\Xi$ in (\ref{kam}) accounts in part for the fact that the vectors $\{\xi_i\}$ in general will depend on $A$, which means that for instance commutators $[\nabla_{\xi_i}, c(\psi_j)]$ will be non-zero, and in part for curvature terms. 

Finally note that as an alternative to the twisted fluctuation in (\ref{dune}) the fermionic Hamilton operator (\ref{kam}) can also be obtained from a Bott-Dirac operator as discussed in \cite{Aastrup:2024ytt}.

\section{Rewriting the fermionic Hamiltonian}

The fermionic Hamilton operator (\ref{kam}) involves operator-valued fields that are one-forms and as such it does not resemble the fermionic sectors we know from for example the standard model of particle physics.  
However, in \cite{Aastrup:2024iqe} we analysed the special case where  there is a triad field on $M$ and found that in this case there exist a change of basis for which the operator (\ref{kam}) can be rewritten in a form of a Dirac Hamiltonian.
To see how this works for the fermionic Hamiltonian (\ref{kam}) let $g$ be a metric on $M$ and let $e$ be an associated triad field, i.e.\footnote{We use standard summation conventions over spatial ($\m,\n,\rho,\ldots$) and Lie-algebra ($a,b,c, \ldots$) indices.}  $g_{\m\n}= e_\m^a e_\n^a$ with $e= e^a_\m dx^\m \sigma^a $ where $\sigma^a$ are the Pauli matrices.
We introduce an orthonormal basis $\{\phi_i\}$ of $L^2(M,\mathfrak{su}(2) \otimes \mathfrak{u}(2))$ 
We denote by $\sigma^a$, $a\in \{1,2,3\}$, the generators of $\mathfrak{su}(2)$ and by $\tau^a$, $a\in \{0,1,2,3\}$, the generators of $\mathfrak{u}(2)$ and write $\phi_i = \phi_i^{ab}\sigma^a \tau^b$. The orthogonality of this basis is with respect to the inner product 
$$
\langle \a  \vert \b \rangle_{\mathfrak{su}(2)\otimes \mathfrak{u}(2)} := \langle e(\a)  \vert e(\b) \rangle_A, \quad \a, \b \in L^2(M,\mathfrak{su}(2)\otimes \mathfrak{u}(2))
$$
where $e(\a)= e_\m^a dx^\m \a^{ab} \tau^b  $. 
Next, we construct an orthonormal basis $\{{\zeta}_i\}$ of $\OO^1(M,\mathfrak{u}(2))$ where 
$
{\zeta}_i =  e^a\phi_i^{ab} \tau^b
$ and express the basis $\{\xi_i\}$ in terms of the new basis
$$
\xi_i = \sum_m {\zeta}_m M_{mi} ,
\qquad 
M_{mi} = \langle {\zeta}_m \vert \xi_i\rangle_A,
$$
which we then use to reformulate the fermionic Hamiltonian 
$H_{\mbox{\tiny fermionic}}$ as \begin{equation}
H_{\mbox{\tiny fermionic}} =  \frac{1}{3!}\int_M \mbox{\small dVol }\mbox{Tr}_{\mathfrak{su}(2)\otimes \mathfrak{u}(2)}\left( \tilde{\Phi}  D^A \tilde{\Phi}^\dagger -  \tilde{\Phi}^\dagger  D^A \tilde{\Phi}  \right) + \Xi,
\label{ssd}
\end{equation}
where
$$
D^A= -\mathrm{i}\sigma^{a} e^\m_a   \left(\nabla^A_\m + \oo_\m \right)  ,\qquad (\oo_{\m})_a^{\;\; b} = e^\n_a \pa_\m e^b_\n,
$$
and
\begin{equation}
\tilde{\Phi}(x) = \sum_{mi} M_{mi}\phi_m (x) \mbox{int}(\xi_i), \quad 
\tilde{\Phi}^\dagger (x) = \sum_{mi} M_{mi} \phi_m(x) \mbox{ext}(\xi_i).
\label{Greenland}
\end{equation}
The operator valued fermionic fields in (\ref{Greenland}) satisfy the relations
\begin{eqnarray}
\{\tilde{\Phi}(x), \tilde{\Phi} (y) \} &=& 0,
\nn\\
\{\tilde{\Phi}^\dagger(x), \tilde{\Phi}^\dagger  (y) \} &=& 0,
\nn\\
\{\tilde{\Phi}^\dagger(x), \tilde{\Phi} (y) \} &=& \sum_{m}  \phi_m(x)\phi_m(y).
\label{foss}
\end{eqnarray}
The integral kernel $\sum_{m}  \phi_m(x)\phi_m(y)$ is proportional to the Dirac delta-function in the limit where the inner product
$\langle \cdot \vert\cdot \rangle_A$ is equal to the $L^2$-norm on $\OO^1(M,\mathfrak{u}(2))$. 
Note that the fermionic fields $(\tilde{\Phi}^\dagger, \tilde{\Phi}) $ are no longer one-forms.

  In total, we see that (\ref{ssd}) is the principal part of the Dirac Hamiltonian for a trival choice of space-time foliation \cite{Paschke} (i.e. lapse and shift fields  $N=1, N^a=0$) and that  (\ref{foss}) are the canonical anti-commutation relations of a quantized fermionic field that takes values in two copies of the Lie-algebra of $SU(2)$. These fermionic fields live on a curved background.

\subsection{Introducing spinors in the Hamiltonian}

The change of basis constructed in the above relies on the map
\begin{equation}
e: TM^\ast \rightarrow C(M)\otimes \mathfrak{su}(2).
\label{shat}
\end{equation}
We can combine this map with the isometry (\ref{chiii}), which can straightforwardly be extended to an embedding
\begin{equation}
{\chi}_{(\varphi_1,\varphi_2) }:L^2 (M,\mathfrak{su}(2)\otimes \mathfrak{u}(2))\to L^2 (M,(S\oplus S)\otimes (S\oplus S)),
\label{chiiii}
\end{equation}
to obtain the diagram
\begin{equation}
\vspace{1mm}\begin{array}{ccc}
\OO^1(M,\mathfrak{u}(2)) & \stackrel{{\chi}_{(\varphi_1,\varphi_2) }}{\looongrightarrow} & \OO^1(M,S\oplus S)
\vspace{2mm}
\\
e \Big\downarrow &&\hspace{-1,6cm}  {\chi}_{(\varphi_1,\varphi_2) } \cdot e \Big\downarrow
\vspace{2mm}
\\
L^2(M,\mathfrak{su}(2)\otimes \mathfrak{u}(2))& \stackrel{{\chi}_{(\varphi_1,\varphi_2) }}{\looongrightarrow} &
 L^2(M,(S\oplus S)\otimes(S\oplus S)).
\end{array}
\vspace{1mm}
\label{ddds}
\end{equation}
Note that (\ref{chiiii}) is an embedding because it involves the map 
$$
{\chi}_{(\varphi_1,\varphi_2) }: \mathfrak{su}(2)\rightarrow S\oplus S.
$$
We can use (\ref{ddds}) to rewrite the quantity $\mbox{Tr}_{\mathfrak{su}(2)\otimes \mathfrak{u}(2)}\left( \phi_m  D^A \phi_n   \right)$ that the fermionic Hamiltonian in (\ref{ssd}) is build on:
\begin{eqnarray}
    \mbox{Tr}_{\mathfrak{su}(2)\otimes \mathfrak{u}(2)}\left( \phi_m  D^A \phi_n   \right) 
    \hspace{-3,4cm}&&
    \nn\\
    &=&
     \mbox{Tr}_{\mathfrak{su}(2)\otimes \mathfrak{u}(2)}\left( \chi^{-1}_{(\varphi_1,\varphi_2) }(\varrho_m)  D^A \chi^{-1}_{(\varphi_1,\varphi_2) }(\varrho_n)   \right)
     \label{horiz}
\end{eqnarray}
where $\{\varrho_i\}$ is a set of orthonormal vectors in $L^2(M,(S\oplus S)\otimes(S\oplus S))$, which are orthonormal with respect to the inner product obtained from $\langle \cdot\vert\cdot\rangle_S$ using (\ref{shat}) 
and which satisfy
\begin{equation}
\varrho_m = \chi_{(\varphi_1,\varphi_2) } (\phi_m).
\label{nablo}
\end{equation}
Next, we rewrite (\ref{horiz}) as
\begin{eqnarray}
    \int_M \mbox{\small dVol }\mbox{Tr}_{\mathfrak{su}(2)\otimes \mathfrak{u}(2)}\left( \phi_m  D^A \phi_n   \right) 
    \hspace{-3cm}&&
    \nn\\
    &=&
     \int_M \mbox{\small dVol }\mbox{Tr}_{\mathfrak{su}(2)\otimes \mathfrak{u}(2)}\left( \chi^{-1}_{(\varphi_1,\varphi_2) }(\varrho_m)  \chi^{-1}_{(\varphi_1,\varphi_2) }(\tilde{D}^A\varrho_n)   \right)
     \nn\\
     &=&
      \int_M \mbox{\small dVol} \left(\varrho_m , \tilde{D}^A\varrho_n\right) + \co(\tau),
\label{ers}
\end{eqnarray}
where $\tilde{D}^A$ is defined by
\begin{equation}
\chi^{-1}_{(\varphi_1,\varphi_2) }(\tilde{D}^A\varrho_n) = D^A\chi^{-1}_{(\varphi_1,\varphi_2) }(\varrho_n),
\label{oerslev}
\end{equation}
and where $(,)$ is the local inner product on $(S\oplus S)\otimes (S\oplus S)$. The parameter $\tau$ in (\ref{ers}) parametrizes the non-locality of $\langle \cdot \vert \cdot \rangle_{\mathfrak{su}(2)\otimes \mathfrak{u}(2)}$, i.e.
$$
\langle \cdot \vert \cdot \rangle_{\mathfrak{su}(2)\otimes \mathfrak{u}(2)}  = \langle \cdot \vert \cdot \rangle_{L^2}
+\co(\tau)
$$
where $\langle \cdot \vert \cdot \rangle_{L^2}$ is an $L^2$-norm on $L^2(M,\mathfrak{su}(2)\otimes \mathfrak{u}(2)) $. 
Note that $\tilde{D}^A$ is only defined on the image  $\chi_{(\varphi_1,\varphi_2)} (L^2(M,\mathfrak{su}(2)\otimes \mathfrak{u}(2)))$ and not on the entire $L^2((S\oplus S) \otimes (S \oplus S))$.

In this way we can rewrite the fermionic Hamiltonian (\ref{kam}) in terms of elements in $L^2(M,(S\oplus S)\otimes (S\oplus S))$
\begin{equation}
H_{\mbox{\tiny fermionic}} = 2  \langle \tilde{\Psi} \vert   \tilde{D}^A\tilde{\Psi}^\dagger \rangle_S - 2  \langle \tilde{\Psi}^\dagger \vert   \tilde{D}^A\tilde{\Psi} \rangle_S  + \co(\tau)+\Xi 
\label{kam22}
\end{equation}
with
\begin{equation}
\tilde{\Psi}(x) = \sum_i \varrho_i(x) \mbox{int}(\xi_i), \quad 
\tilde{\Psi}^\dagger (x) = \sum_i \varrho_i(x) \mbox{ext}(\xi_i),
\label{llutz2}
\end{equation}
and 
$$
\tilde{D}^A =
\left(
\begin{array}{cc}
-\mathrm{i} \sigma^a e_a^\m\left(  \nabla^A_\m + \oo_\m\right) & 0 \\
0 & -\mathrm{i} \sigma^a e_a^\m\left(  \nabla^A_\m + \oo_\m\right)
\end{array}
\right)
.
$$
Again note that the identity (\ref{kam22}) only involves elements in 
$$\chi_{(\varphi_1,\varphi_2)} (L^2(M,\mathfrak{su}(2)\otimes \mathfrak{u}(2))).$$
On the complement of $\chi_{(\varphi_1,\varphi_2)} (L^2(M,\mathfrak{su}(2)\otimes \mathfrak{u}(2)))$ in 
$L^2((S\oplus S) \otimes (S \oplus S))$ the expression (\ref{kam22})  is zero.

Let us finally note that it is not strictly necessary to apply the map 
$\chi_{(\varphi_1,\varphi_2) }$ twice as we have done it here. As we start with the space $\OO^{1}(M,\mathfrak{u}(2))$ we can use the map $\chi_{(\varphi_1,\varphi_2) }$ to embed the cotangent space of $M$ into $S\oplus S$ and/or map the Lie-algebra $\mathfrak{u}(2)$ into $S\oplus S$. The first mapping is essential for our analysis whereas the second is not.

\section{Hilbert space and semi-classical states}

Let us now introduce the Hilbert space in which $D_\cf$ acts:
$$
\ch = \left( L^2(\cf) \oplus L^2(\cf)  \right)\otimes \bigwedge^* \OO^1(M,\mathfrak{u}(2))
$$
where $L^2(\cf)$ 
is the Hilbert space over $\cf$ constructed in \cite{Aastrup:2020jcf} and \cite{Aastrup:2023jfw}. In \cite{Aastrup:2017vrm} we proved that a representation of the $\mathbf{HD}$-algebra exists in the one-particle sector of $\ch$ and in \cite{Aastrup:2023jfw} we extended this representation to a fermionic Fock space build over $\OO^1(M,\mathfrak{g})$, where $\mathfrak{g}$ was the Lie-algebra of any compact Lie-group\footnote{It is important to note that these representations mentioned here either break gauge-invariance or presumes the resolution of the Gribov ambiguity \cite{Gribov:1977wm}. For details see \cite{Aastrup:2023jfw}.}. In the following we will show how this representation for $\mathfrak{g}=\mathfrak{u}(2)$ can be extended to $\ch$.  

We use the map (\ref{alien}) to identify $\bigwedge^* \OO^1(M,\mathfrak{u}(2))$ with $\bigwedge^* \OO^1(M,S \oplus S)$
 and focus first on the one particle subspace $L^2(\cf, \OO^1(M,S\oplus S))$. We let $A$ be a connection in $\cf$ and $e^X$ a flow. We need an action $e^X_A$ of $e^X$ on $\OO^1(M,S\oplus S)$, which means that we need to describe how $e^X_A$ moves an element in $\OO^1(M,S\oplus S)$. 
To this end we denote by $\{\psi_i\}$ a set of orthonormal vectors in $\OO^1(M,S\oplus S)$ given by
$$
\psi_i = \chi_{(\varphi_1,\varphi_2) } (\xi_i),
$$ 
and write for the spin-part
 \begin{equation}
 (e^X_A \psi_i )(\exp_1(X)(x))=
  (\hbox{Hol}(\gamma , A) \psi_i (x) ),
 \label{sveri}
 \end{equation}
with $\gamma(t)=\exp (X)(x)$, where we use a four-dimensional irreducible representation of $U(2)$.  Concerning the one-forms, then we can move them with $e^{-X}$ since it is a diffeomorphism (note that one-form transforms contra-variantly). 

Note that the representation (\ref{sveri}) is not the same as the one we worked with in \cite{Aastrup:2023jfw}, where we used for the $\mathfrak{su}(2)$-factor the adjoint action of $SU(2)$ on $\mathfrak{su}(2)$. Since we now have the spin-bundle $S$ we cannot use an adjoint representation, but as we work with $\mathfrak{u}(2)$ this is also not necessary.  

 Another important point, which we also mentioned in \cite{Aastrup:2019iyp}, is that this representation is not unitary, not even after multiplying with $(\Delta \exp_1) (x)$ as in (\ref{chopin1}). The reason is that the diffeomorphism $e^X$ in general does not act unitarily on one-forms. However, if the manifold $M$ is compact then $e^X_A$ is a bounded operator on $\OO^1(M,S\oplus S)$. 

  Finally, to obtain a representation on all of $\ch$ we extend (\ref{sveri}) multiplicatively to all of $\bigwedge^* \OO^1(M,S \oplus S) $. Note that $e^X_A$ will not be a bounded operator unless $M$ is compact and we restrict ourselves to states with a finite number of particles.

\subsection{The emergence of an almost-commutative algebra}

We will now consider the case where we have a state $\Psi_{A_0}(A)$ in $L^2(\cf)$ that is sharply localised in a single point $A_0$, i.e. if $F(A)\in \mathscr{F} (\ca , \mathbb{B} (L^2(M, V) ))$ then
$$
F(A)\Psi_{A_0}(A) = F(A_0)\Psi_{A_0}(A) + \co(\tau)
$$
for some parameter $\tau$.
In this section we will show that the $\mathbf{HD}$-algebra in this case reduces to an almost-commutative algebra.

To see this let us first consider closed flows, i.e. flows where $\exp_1(X)$ is the identity on $M$. We will only consider the action of $e^X$ on the one particle space, i.e. the action on $ \Omega^1 (M,S\oplus S)$. Since $\Psi_{A_0}(A)$ is sharply localized in $A_0$ to lowest order in $\tau$ we have 
$$  e^X (\Phi )(m) =Hol (\gamma_m ,A_0) \Phi (m) , $$
where $Hol (\gamma_m ,A_0)$ acts on the $S\oplus S$ component of $\Phi \in \Omega^1 (M,S\oplus S)$, and where $\gamma_m (t)=\exp_t (X) (m) $. Thus to lowest order in $\tau$ $e^X$ acts as a function over $M$ with values in the unitaries of  $End (S\oplus S)$. If the connection $A_0$ is degenerate these unitaries are all trivial, but if $A_0$ is not degenerate the $\mathbf{HD}$-algebra will in this limit generate a lot of nontrivial unitary functions, and since the $\mathbf{HD}$-algebra also contains the sum of flows we get in general a rich sub-algebra of smooth functions with values in  $End (S\oplus S)$.  

Let us look at what $A_0$ must fulfill in order for this sub-algebra to be dense in $C^\infty (M,End(S\oplus S))$. According to Stone-Weierstra{\ss} it suffices to require
\begin{enumerate}
\item The sub-algebra must be a unital $*$-algebra. This follows, since the $\mathbf{HD}$-algebra is a unital $*$-algebra.
\item The sub-algebra must separate the pure states. A pure state is given by a point $m\in M$ together with a normalized vector in $S\oplus S$ over $m$. So the requirement is that for two given points $m_1$ and $m_2$ and vectors $v_1$ and $v_2$,  $\| v_1\|=\|v_2\|=1$,  we must find an element $f$ in the sub-algebra with 
$$ \langle v_1 | f(m_1)|v_1\rangle \not=  \langle v_2 | f(m_2)|v_2\rangle .$$
\end{enumerate}
Note that if we consider the diagonal representation of $U(2)$ on $S\oplus S$ the second condition is of course never fulfilled. But a generic $A_0$ {should} fulfill the second condition on each of the copies of $S$.  Also if we take the 4-dimensional irreducible representation of $U(2)$ acting on $S\oplus S$ the second condition { should} for a generic connection be fulfilled. Especially since we can also consider products $e^Xe^Y$, where $X$ and $Y$ do not have to describe closed flows, but only the composition of the two flows has to be closed.

To conclude, we find that the part of the $\mathbf{HD}$-algebra that consist of closed loops will give rise to a sub-algebra of the almost-commutative algebra
$$
C^\infty (M)\otimes M_4(\mathbb{C})
$$
 in a semi-classical limit, and that if the connection, in which we localize, is sufficient non-degenerate, then the $\mathbf{HD}$-algebra will generate the entire $
C^\infty (M)\otimes M_4(\mathbb{C})
$.

Note that since we are representing the  $\mathbf{HD}$-algebra  on $L^2 (\cf )\oplus L^2 (\cf )  $ we can choose a different algebra than just one copy of the $\mathbf{HD}$-algebra.  For example we can choose $M_2 (\mathbb{C})\otimes \mathbf{HD}$ to act on $L^2 (\cf )\oplus L^2 (\cf )  $ instead of the diagonal representation of the $\mathbf{HD}$-algebra. In this case the algebra will reduce to $
C^\infty (M)\otimes M_8(\mathbb{C})
$  in the semi-classical limit. Another option is to choose the algebra $\mathbf{HD}\oplus \mathbf{HD}$, which in the semi-classical limit would give us 
$$
C^\infty (M)\otimes (M_4(\mathbb{C})\oplus M_4(\mathbb{C}))
$$ 
provided that we choose the semi-classical state to be $\Psi_{A_1}\oplus \Psi_{A_2}$, where $A_1$ and $A_2$ are classical points in $\cf$. However, in this case the semi-classical state is no-longer a pure state and is also not centered around one classical configuration.

Let us finally stress that the specific form of the finite factor of the emerging almost-commutative algebra depends on the choice of representation of the $\mathbf{HD}$-algebra.

\subsection{The structure of an almost-commutative spectral triple}

Let us now look at the interaction between the almost commutative algebra and the Hamilton operator $H_{\mbox{\tiny fermionic}}$. We will look at it on one particle states, i.e a function of the form $\Psi_{A_0}\otimes \Phi$, with $\Phi \in \Omega^1 (M,S\oplus S)$.  The action of the almost commutative algebra on $\Psi_{A_0}\otimes \Phi$ is basically just by fiber-wise matrix multiplication over $M$ on the $\Phi$. In addition to this $H_{\mbox{\tiny fermionic}}$ acts as $\tilde{D}^{A_0}$ on elements in $\Omega^1 (M,S\oplus S)$. Thus the commutator between the almost commutative algebra and $H_{\mbox{\tiny fermionic}}$ gives us the standard commutator between matrix valued functions on $M$ and the Dirac type operator $\tilde{D}^{A_0}$.

To summarize, we find the general structure of a Hamilton formulation \cite{Paschke} of an almost-commutative spectral triple emerges from our construction in a semi-classical limit:
$$
(\cb_F,\tilde{D}^{A_0}, L^2(M,(S\oplus S)\otimes (S\oplus S)) )
$$
where $\cb_F$ is a sub-algebra of $C^\infty (M)\otimes M_4(\mathbb{C})$ and where
the operator $\tilde{D}^{A_0}$ is the spatial part of a four-dimensional Dirac operator. The Hilbert space 
$L^2(M,(S\oplus S)\otimes (S\oplus S))$ has a double-fermionic structure where the matrix part of $\cb_F$ algebra acts on the second factor. The operator
$\tilde{D}^{A_0}$ acts on both spinor-factors in the Hilbert space. If we write $\tilde{D}^{A_0}$ as 
$$
\tilde{D}^{A_0} = \tilde{D}_M + \mathbf{1}_2\otimes\sigma^a e_a^\m (A_0)_\m,
$$
then $\tilde{D}_M$ acts in the first spinor-factor in the Hilbert space, while $\mathbf{1}_2\otimes\sigma^a e_a^\m (A_0)_\m$ acts in both factors, with the matrices $\mathbf{1}_2\otimes\sigma^a$ in the first factor and with the gauge field in the second factor.

\section{Discussion}

The picture that emerges from this paper is that of a fundamental theory that is based on the dynamical geometry of a configuration space, and which on the one hand gives rise to a $U(2)$ Yang-Mills-Dirac quantum field theory and on the other hand renders the general structure of a Hamiltonian formulation of an almost-commutative spectral triple in a semi-classical limit. The latter shows some similarities to the almost-commutative spectral triple identified by Chamseddine and Connes in the standard model of particle physics. 

Perhaps the most pressing question that these results raise is one of  interpretation. Specifically, in the semi-classical limit the part of the emerging spatial Dirac type operator\footnote{As we have already mentioned the operators $\tilde{D}^A$ and $D^A$ are strictly speaking not Dirac operators but rather three-dimensional projections of four-dimension Dirac operators. This is what one would expect to find in a Hamilton formulation of four-dimensional spectral triple \cite{Paschke}.} that interacts with the finite factor of the almost-commutative algebra comes from the gauge field over which the semi-classical state is peaked. This unusual setup suggests that the Yang-Mills quantum field theory that emerges from the Dirac operator on the configuration space cannot be directly equated with the bosonic sector of the standard model but should instead be interpreted in terms of a more fundamental theory, whose semi-classical limit gives rise to the aforementioned almost-commutative structures. This more fundamental theory will then give rise to quantum corrections and the question will therefore be if these may account for the quantum corrections found in the standard model.

To fully answer the question, whether our construction is connected to the spectral formulation of the standard model, we would first have to convert our construction to a path integral formulation and check whether the spectral formulation of the standard model does indeed emerge in a semi-classical limit. Such an analysis has not been carried out. Secondly, we would probably need to see the full toolbox of noncommutative geometry emerge from our construction in that semi-classical limit. The point is that if our construction is to be a credible candidate for a fundamental theory we cannot simply impose the axioms, on which Chamseddine and Connes' work is based, on our construction; rather, they too should be emergent. The question is to what extent this is possible.  

 Apart from the emergence of the aforementioned almost-commutative structures a further reason to believe that a connection to the spectral standard model is possible is that our construction comes with a double fermionic structure that is similar to the fermionic structure found in the spectral formulation of the standard model. In the latter case this lead to the so-called fermion doubling problem \cite{Lizzi:1996vr}, which, in turn, is related to the KO-dimension of the noncommutative geometry behind the standard model. It is an interesting question what the KO-dimension of our construction is, but in order to determine that we need a real structure. In \cite{Aastrup:2024xxl} we proposed a candidate for a real structure, but that candidate did not fully adhere to the axioms of noncommutative geometry, a fact that may indicate that the Dirac operator used in that paper should be changed. Indeed, based on our most recent papers \cite{Aastrup:2023jfw,Aastrup:2024iqe,Aastrup:2024xxl} it is clear that there is room for variation when it comes to the exact form of the Dirac operator on $\cf$ and the representation of the $\mathbf{HD}$-algebra.

Another key point worth mentioning is that several of the result obtained in this paper depends on the emergence of a triad field in the semi-classical limit. We have previously argued that a metric field on $M$ should emerge from our construction in a semi-classical limit \cite{Aastrup:2023jfw}, but more analysis is required to determine whether this is indeed the case. 
Similarly, a key ingredient in our analysis is the map $\chi_{(\varphi_1,\varphi_2) }$ between the Lie-algebra of $U(2)$ and the space $S\oplus S$. This map, which introduces spinors into our construction, is unusual since it potentially maps spin-one objects into spin-half objects. We have previously found that the metric on $\cf$ is independent on the choice of the two spinors $(\varphi_1,\varphi_2)$ under certain conditions, but more analysis is required to determine to what extend the whole construction is independent of this choice.

\vspace{1cm}
\noindent{\bf\large Acknowledgements}\\

\noindent

$   $
JMG would like to express his gratitude to entrepreneur Kasper Bloch Gevaldig for his generous financial support. JMG is also indepted to entrepreneur Jeff Cordova as well as the following list of sponsors for generous support:   
 Frank Jumppanen Andersen,
 Miroslav Bajtos,
 Danny Birkmose,
 Simon Chislett,
 Jos van Egmond,
 Trevor Elkington,
 Claus Hansen,
 David Hershberger,
 JaeGyu Kim,
 Simon Kitson,
 Hans-J\o rgen Mogensen,
 Stephan M{\"u}hlstrasser,
 Paul Sharma,
 Rolf Sleimann,
 Ben Tesch,
 Adam Tombleson,
 Jeppe Trautner,
 Vladimir Zakharov,
and the company
 Providential Stuff LLC. JMG would also like to express his gratitude to the Institute of Analysis at the Gottfried Wilhelm Leibniz University in Hannover, Germany, for kind hospitality during numerous visits.

\end{document}